\documentclass[prb,twocolumn,amsmath,amssymb]{revtex4}

\usepackage{color}
\usepackage{graphicx}
\usepackage{dcolumn}
\usepackage{bm}
\def\be{\begin{equation}}
\def\ee{\end{equation}}
\def\bd{\begin{displaymath}}
\def\ed{\end{displaymath}}
\def\-{\phantom{-}}

\begin{document}

\title{Electric field effect on superconductivity at complex oxide interfaces}

\author{Jason T. Haraldsen$^{1,2}$}
\author{Peter W\"olfle$^{3}$}
\author{Alexander V. Balatsky$^{1,2}$}
\affiliation{$^{1}$Theoretical Division, Los Alamos National Laboratory, Los Alamos, NM 87545, USA}
\affiliation{$^{2}$Center for Integrated Nanotechnologies, Los Alamos National Laboratory, Los Alamos, NM 87545, USA}
\affiliation{$^{3}$Institute for Condensed Matter Theory and Institute for Nanotechnology, Karlsruhe Institute of Technology, D-76128 Karlsruhe, Germany}

\begin{abstract}

We examine the enhancement of the interfacial superconductivity between LaAlO$_{3}$ and SrTiO$_{3}$ by an effective electric field. Through the breaking of inversion symmetry at the interface, we show that a term coupling the superfluid density and an electric field can augment the superconductivity transition temperature. Microscopically, we show that an electric field can also produce changes in the carrier density by relating the measured capacitance to the density of states. Through the electron-phonon induced interaction in bulk SrTiO$_{3}$, we estimate the transition temperature.

\end{abstract}

\maketitle

\section{Introduction}

Superconductivity is typically thought to be produced by the pairing of electrons below a critical temperature $T_c$.\cite{bard:57} The overall pairing interaction is dependent on a subtle balance of the repulsive Coulomb interaction and any attractive interaction supplied by the exchange of collective excitations in the system.\cite{bard:57,park:69}  That balance depends, in a complex way, on the number of carriers within the material.\cite{park:69,ande:97,kett:99}  This effect becomes all the more important in the case of spatial inhomogeneity, e.g. at an interface, where the spatial distribution of the carrier density will be important.\cite{okam:04,taka:06} Therefore, it is of great interest to have a tuning parameter like an external electric field producing a change to the carrier distribution and effectively maximizing the transition temperature $T_c$.

\begin{figure}
\includegraphics[width=3.5in]{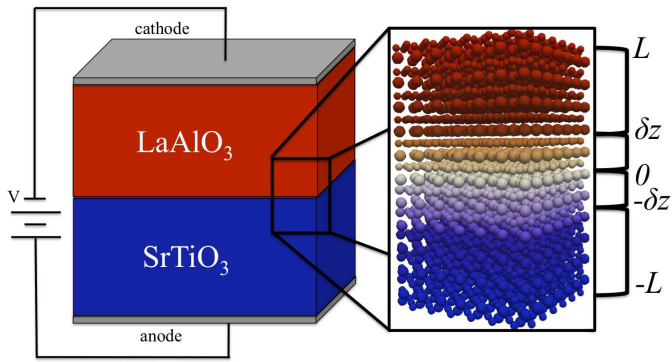}
\caption{An illustration of LaAlO$_3$/SrTiO$_3$ interface, which details the interface region. The interfacial (lighter color) area is a small region from $-\delta z$ to $\delta z$. The bulk (darker color) region is that volume that is essentially unaffected by the interface interactions. For the purpose of this study, we examine the effect of an applied electric field produced throughout the sample. Therefore, all calculations are to mimic this general setup.}
\label{LAOSTO}
\end{figure}

In recent years, oxide heterostructures have attracted much attention due to a rich spectrum of unexpected behavior discovered at interfaces.\cite{hebe:09,hwan:06} Specifically, the emergence of a 2-D electron gas, that becomes superconducting at around 200 mK, at the epitaxially grown interface of the insulator SrTiO$_3$ (STO) and a few atomic layers of LaAlO$_3$ (LAO) (illustrated in Fig. \ref{LAOSTO}) has sparked an intense search for the mechanism of superconductivity as well as for improved materials.\cite{hebe:09,hwan:06,ohto:02,ohto:04,reyr:07} Studies investigating the electron mobility have demonstrated that intrinsic doping may be key to understanding the underlying mechanisms. \cite{siem:07} While the superconductivity occurs at much lower temperatures than in high-T$_c$ superconductors, this finding shows the importance of investigating interfacial physics in heterostructures. Even though there are limitations due to lattice matching and strain within the materials,\cite{naka:06,waru:09} the coupling between these materials may provide information that is critical to the understanding of these heterostructures. 

Further studies of the LAO/STO system have shown that $T_c$ can be increased by an applied electric field.\cite{cavi:08,bell:09} Bulk STO has been shown to become superconducting with doping by oxygen vacancies and by niobium (Nb) at about the same temperature as observed in the LAO/STO system.\cite{scho:64} Furthermore, STO has been found to exhibit interfacial superconductivity when brought in connect with an oxide gel in the presence of an external electric field.\cite{ueno:08} These studies show that $T_c$ can be induced and/or enhanced by the creation of carriers pulled from a source (i.e. LAO, oxide gel, or doping).\cite{cavi:08,bell:09,ueno:08} 

The nature of the superconducting state at the interface is a subject of ongoing discussion. One possibility is that the superconducting state is conventional electron-phonon driven, but there is also a growing list of possible unconventional states including magnetism driven ones.\cite{koer:05,step:11} Indeed recent data from scanned probes point to a complicated spatial pattern of coexisting superconductivity and magnetism at the interfaces.\cite{bert:11} Early reports on ferromagnetic order coexisting with superconductivity \cite{brink:07} have been independently confirmed.\cite{li:11}  For an assessment of the present situation see Ref. [\onlinecite{millis:11}].

For the following, it will be important to know the thickness of the mobile electrons layer at the interface. Early estimates of the 2-D density of mobile carriers have yielded values much below the theoretical prediction of 0.5 electrons per unit cell.\cite{thiel:06} Apparently, most of the added charge is not mobile and is trapped within a few unit cells of the surface, as observed by high-energy photoemission experiments,\cite{sing:09} while the mobile electrons may cover a region of thickness 10 nm or more. 

The recent experiments \cite{bert:11} suggest that the spatial separation of superconducting and ferromagnetic regions may allow spin singlet pairing to coexist with ferromagnetic domains. The spatial extension of  the density of mobile carriers normal to the interface  over several tens or even hundreds of unit cells speaks against a prominent role of the on-site Coulomb interaction and therefore a pairing mechanism induced by magnetic fluctuations. All of this points to a conventional s-wave superconducting state. 

In this study, we investigate i) the effect of an electric field applied externally on the 2-D and 3-D carrier density. We  calculate the variation of the 2-D carrier density, capacitance, and the 3-D carrier density depth profile with increasing electric potential for LAO/STO heterostructures. From changes of the overall carrier density at the interface, we deduce that the nonlinear capacitance of the layer at the interface implies an energy dependent Density of States (DOS); ii) using modifications of charge carrier density and DOS as a tunable parameter, we calculate changes in superconducting transition temperature $T_c$.  We assume the pairing to be conventional s-wave pairing as identified in detailed studies of doped bulk STO.\cite{scho:64}  Using this model assumption, we show that the $T_c$ for the band insulator STO can be modified by changes in charge density, and we find that $T_c$ does exhibit a maximum at a certain carrier density determined by the peak in the capacitance. We show a plausible connection between the increase in carrier density and $T_c$ due to an applied electric field. This observation could lead to a possible mechanism for a tunable functional response: dramatic changes in the superconducting properties with the application of external fields. 

 \begin{figure}
\includegraphics[width=3.5in]{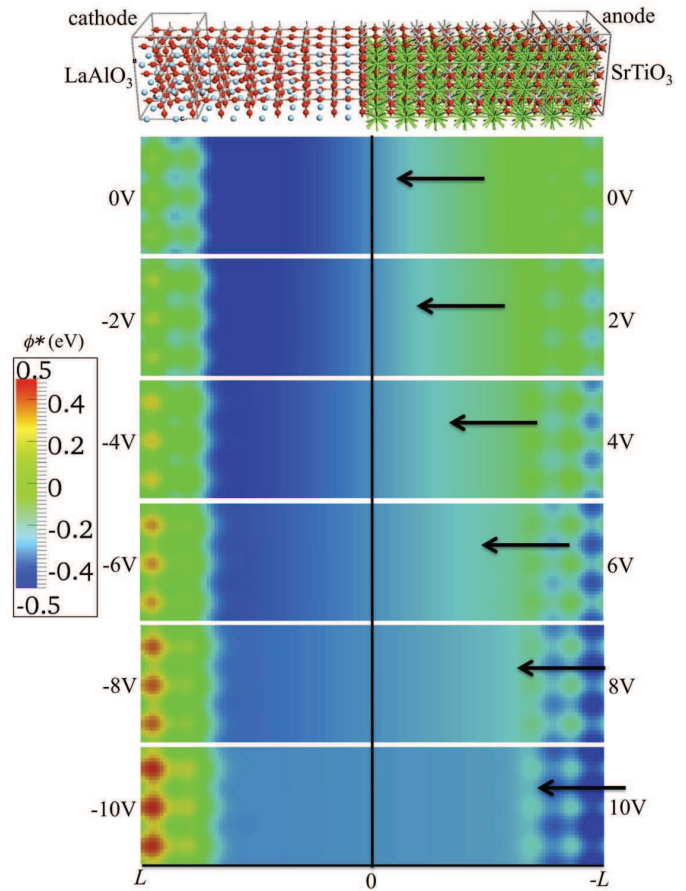}
\caption{Model calculation for the shifting electrostatic difference potential (EDP) due to an applied electric potentials ranging from 0 V to 10 V using a semi-empirical, extended H\"uckel model for the LAO/STO interface.\cite{sole:02,quantum} The model shows the diffusion of the potential (approximately 1 unit cell per 2V) from the LAO region into the STO region (the arrows are to help guide the eye). In the simulation, the last two unit cells are used as the electrodes (LAO - cathode and STO - anode) producing an overall electric potential over the sample region.}
\label{EDP}
\end{figure}

\section{Field Effect on the Ginzburg-Landau Free Energy}

The breaking of inversion symmetry at the interface of heterostructures introduces the interface normal vector as a preferred direction of the system.\cite{hara:11} The normal component of the electric field $F_z$ is therefore compatible with the symmetries of the free energy, allowing a coupling term of electric field and superconducting density, $F_z |\psi|^2$, with the superconducting order parameter $\psi$. This is a simple consequence of screening effect on the charge density within the superconductor.\cite{linearnote} The charge distribution at the interface would be dependent on the distance from interface, typically on the order of nm. On the other hand the relevant superconducting length scale is set by a fairly large coherence length $\xi/2$, typically on the scale of tens to hundreds of nm.\cite{ande:97} Therefore, we may use a Ginzburg-Landau (GL) functional in our approach when we address the coupling between SC and electric degrees of freedom.\cite{kett:99,park:69} 

Since the interface has the additional constraints of boundary conditions and strain (as shown in Fig. \ref{LAOSTO}), the standard heterostructure free energy  can be broken into two main components consisting of bulk ($B$) and interfacial ($I$) regions. The effect of the interface decays off as one moves into the bulk. In the case of the LAO/STO interface, the total free energy $FE$ is then given by
\be
\begin{array}{r}
\displaystyle FE_{{\rm tot}} = \int_{-L}^{-\delta z} F(z)_{B^{{\rm STO}}} dz + \int_{-\delta z}^{0} F(z)_{I^{{\rm STO}}} dz\\ \\ \displaystyle + \int_{0}^{\delta z} F(z)_{I^{{\rm LAO}}} dz+ \int_{\delta z}^{L} F(z)_{B^{{\rm LAO}}} dz,
\end{array}
\label{G-L} \ee
where the depth-dependent free energy for the bulk and interface are the given by
\be
FE(z)_{B} = \alpha_{B} |\psi(z)_{B}|^2 + \frac{\beta_{B}}{2} |\psi(z)_{B}|^4
 \ee
and
\be
\begin{array}{c}
FE(z)_{I} = \alpha_{I} |\psi(z)_{I}|^2 + \frac{\beta_{I}}{2} |\psi(z)_{I}|^4 + g
|\bigtriangledown \psi(z)_{I}|^2 \\ \\ + \lambda F_{z} |\psi(z)_{I}|^2 + \gamma \eta^2|\psi(z)_{I}|^2,
\end{array}
\label{GL-z} \ee
where $\alpha_{B,I}$ and $\beta_{B,I}$ are the standard GL energy parameters. In the standard GL formalism, the parameter  $\alpha_{B}$ = $a_{B}(T-T_{c}^{0})$, with $a>$ 0, determines the normal ($\alpha_{B}>0$) and superconducting states ($\alpha_{B}<0$), while $\beta_{B}$ is always positive and helps stabilize the ground state.

\begin{figure}
\includegraphics[width=3.5in]{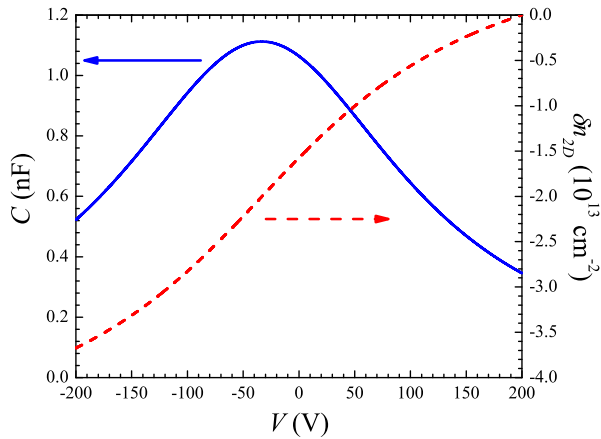}
\caption{Simulated capacitance and 2-D carrier density $n_{2d}$ as a function of electric potential for the LAO/STO interface in Ref. [\onlinecite{cavi:08}]. It should be noted that the change in curvature denotes an effective change in electron or hole doping.}
\label{candn2d}
\end{figure}

For the interface region, the gradient and strain $\eta$ terms are produced by the interface and are typically positive, where the gradient produces the boundary condition for the interface and can be assumed to be in the small wavevector limit. For the field interaction, we use ${\bf F} = (0,0,F_z)$. The coefficient $\lambda$ will in general depend on the electric field strength. Therefore, the local carrier density can increase or decrease depending on positive or negative $F_z$, respectively. It should be noted that while the electric field is global, only the superconductivity at the interface will increase due to the accumulation of carriers from the carrier source. This may be different for a high temperature superconductor, but given that STO is an insulator, the distribution of carriers is lessened.

Focusing on the effects of the additional quadratic terms, it can be easily shown that the standard GL definitions are changed. The inclusion of the electric-field interaction and strain terms induces a shift in $T_c$. Concentrating on the $\psi^2$ terms, we re-write the free energy as
\be
\begin{array}{l}
FE(z)_{I}^{\psi^2} = \alpha_{I} |\psi(z)_{I}|^2 + \lambda F_{z} |\psi(z)_{I}|^2+ \gamma \eta^2|\psi(z)_{I}|^2 \\\\ ~ ~ ~ ~ ~ = {\tilde \alpha} |\psi(z)_{I}|^2.
\end{array}
\ee
where $\tilde \alpha$ = $\alpha_{I} + \lambda F_z + \gamma \eta^2$ = $\tilde a(T-T_c^0)$ and describes the superconducting state, when $\tilde \alpha<$ 0 \cite{ande:97,kett:99,park:69}.  Therefore, the shifted critical temperature is given by
\be
T_c =  T_{c}^{0} - \left(\frac { \lambda_c F_z + (\gamma \eta^2)_c}{\tilde a}\right)
\ee
where $\lambda_c$ and $(\gamma \eta^2)_c$  are taken at $T=T_{c}^{0}$. From this we conclude that the electric field has to compete with the lattice strain. Even a large lattice strain may be overcompensated by a reasonably strong electric field, which therefore can still effect the superconducting order parameter. However, if the lattice matching is close, then the strain can be assumed to have a negligible effect on the superconducting state. Recent measurements indicate the various degrees of lattice mismatch may even produce an increase in the superconducting $T_c$.\cite{serq:01}

Thus, the addition of extra quadratic terms to the free energy modifies the critical temperature. While the strain terms usually produce a negative shift of $T_c$, a suitably directed electric field could  gives rise to a positive shift. This may be interpreted as being due to an increase in the carrier density of the superconductor. While this is shown here only in the phenomenological GL model, we will now turn to a modeling of the charge density induced by an applied electric field and show how that is correlated with $T_c$.

\section{Semi-emprical modeling of the electrostatic difference potential}

To examine the change in the electron density at an LAO/STO interface, we calculate the electrostatic difference potential (EDP) $\phi^*$ using a semi-empirical, extended H\"uckel model. The EDP is described as the difference between the electrostatic potential $\phi$ of the self-consistent valence charge density and the electrostatic potential from atomic valence densities.\cite{sole:02,quantum} Therefore, the EDP is determined by solving the Poisson equation $\nabla^2 \phi^*$=$-\rho^*/\epsilon_r$, where is $\rho^*$ is the electron difference density.

Figure \ref{EDP} shows the calculated EDP for various applied electric potentials. This calculation is produced by building a three dimensional interface of LAO and STO that is 3 x 3 x 8 unit cells for each side. The electric potential is produced by simulating a device where positive and negative electrodes are place on the ends of the interface. For this simulation, the LAO electrode is negative and the STO electrode is positive. As the electric potential is increased from 0V to 10V, the EDP is observed to diffuse from LAO into STO. This details the shifting of the electron charge density through the interface. To gain a more detailed view of the change in the charge and carrier densities, we examine the general electrostatics of the interface and compare to known experimental data.

\section{Carrier Density and Capacitance}

\begin{figure}[b]
\includegraphics[width=3.75in]{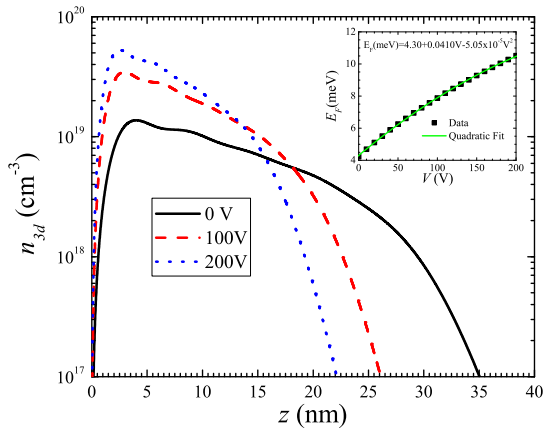}
\caption{Simulated 3-D carrier density $n_{3d}$ as a function of depth for various electric potentials (0V - solid black, 100V - dashed red, and 200 - dotted blue) for the LAO/STO interface. This describes the increase in the overall carrier density at the interface as the electric potential is increased. As this occurs, the relative interfacial region reduces in depth. The inset shows the effective Fermi energy $E_F$ as a function of electric potential. A fitting of this data to a quadratic polynomial provides a general functional form.}
\label{n3d}
\end{figure}

The basic connection of the charge density and an applied electric field is given by integrating Gauss's law for a dielectric medium, keeping the nonlinear dependence of the dielectric function $\epsilon(F)=\epsilon_0\epsilon_r(F)$ on electric field. This function may be determined through the measured capacitance at the LAO/STO interface.
The change in the 2-D carrier density $n_{2d}$ in a semiconducting channel, induced by an electric field is then given by
\be
n_{2d}(F) = \frac{2\epsilon_0}{e} \int_0^{F}\epsilon_r(F) dF
\label{n2d}
\ee
where $F$ is the effective electric field within the sample, $\epsilon_0$ is the permittivity of free space, $\epsilon_r(F)$ is the electric field dependent relative permittivity, and $e$ is the basic unit charge. The field dependence of the relative permittivity for STO has been shown to be $\epsilon_r(F)^{STO} = 1/A\left(1+\frac{B}{A}F\right)$, where $A$ and $B$ are temperature dependent variables determined experimentally (4.097x10$^{-5}$ and 4.907x10$^{-10}$ m/V for STO at 4.2K, respectively). This however produces a carrier density which has a singularity and is discontinuous at $F = -A/B$. The singularity can be removed by expanding out the denominator to a quadratic term, producing a continuous function for all $F$,
\be
\epsilon_r(F)^{STO} = \frac{1}{A\left(C_1+C_2\frac{B}{A}F+C_3\left(\frac{B}{A}\right)^2F^2\right)}
\ee
where $C_i$ ($i$ = 1, 2, and 3) are determined from experimental fits. From this, we find that
\be
n_{2d}(F) =\frac{4\epsilon_0\left( \tan^{-1} \left( {\frac { \left( 2C_3BF+{ C_2}
\,A \right) }{A\sqrt {C_4}}} \right) -  \tan^{-1} \left( {\frac { { C_2} }{\sqrt {C_4}}} \right)\right)}{eB\sqrt {C_4 }} 
\label{n2d}
\ee
where $C_4=4\,{ C_1}\,{C_3}-{{ C_2}}^{2}$. 

To examine the modulation of the carrier density along the LAO/STO interface, we compare to data provided by Ref. [\onlinecite{cavi:08}]. Here, a gate voltage is used to obtain an electric field, where $F$=$F_0 + V/d$, $V$ is the electric potential and $d$ is the sample thickness (0.5mm). The 2-D carrier density is also related to the capacitance $C(V)$ by
\be
C(V) = eS\left(\frac{\partial \delta n_{2d}}{\partial V}\right),
\ee
where $S$ is the area of capacitor and $\delta n_{2d}$ = $n_{2d}(V)-n_{2d}(200V)$ (measuring V in Volts). The cross-sectional area of the capacitor is determined by adjusting the simulated peak capacitance to match the the observed capacitance. Using this method, the cross-sectional area is determined to be about 5 mm$^2$.
Therefore, the capacitance will be
\be
C(V) = \frac{2\epsilon_0S}{Ad\left(C_1+C_2\frac{B}{A}(F_0+\frac{V}{d})+C_3\left(\frac{B}{A}\right)^2\left(F_0+\frac{V}{d}\right)^2\right)}.
\ee
Comparing to the data in Ref [\onlinecite{cavi:08}] and adjusting only the $C_i$ parameters, Figure \ref{candn2d} shows the capacitance and 2D carrier density as a function of gate voltage for STO. Through fitting of the $\delta n_{2d}$ data from Ref.[\onlinecite{cavi:08}], the extended parameters ($C_1$, $C_2$, and $C_3$) are determined to be 4.25, -0.37, and 0.29, respectively with a value of $F_0$ = 1.2x10$^{5}$ V/m. Here, we find that the change in the 2-D carrier density shifts from electron to hole doping around 30 V. This produces the corresponding peak in the capacitance. Further analysis in Section V will correlate the capacitance to the change in $T_c$.

To understand the accumulation of charge at the LAO/STO interface as a gate voltage is applied, we examine the 3D carrier density as function of depth. The depth profile of the charge accumulation in STO is given by summing over the density profiles $|\xi_{i,j}(z)|^2$ of all occupied states
\be
n_{3d}(F,z)  = \sum_{j=l,h}\left(\frac{g_j m_j^*}{2\pi \hbar^2}\sum_i(\Delta E_{i,j})\Theta(\Delta E_{i,j})|\xi_{i,j}(z)|^2\right),
\ee
where $\Delta E_{i,j}$ = $E_F - E_{i,j}$, and the sum over $j=l,h$ is over the light and heavy electronic bands in STO. The inner sum is over the occupied states labelled $i,j$ for the energy of eigenstate $E_{i,j}$ and  Fermi energy $E_F$, where the $\Theta$ function assures that all states $E_{i,j}<E_F$ are included.  $m_j^*$ is the effective mass ($m_l^*$ = 1.2$m_0$ and $m_h^*$ = 4.8$m_0$), $m_0$ is the electron mass, and $g_j=1,2$ is the band degeneracy.\cite{ster:72,bell:09} The confining potential is well represented by a triangular shape and the corresponding eigenstates are therefore normalized Airy functions: 
\be
\xi_{i,j}(z)=n_c{\rm Ai}\left[\alpha_j \left(z- \frac{E_{i,j}}{eF}\right)\right],
\ee
where $n_c$ is the normalization constant in units of (length)$^{-1}$, $E_{i,j} = \frac{eF}{\alpha_j}\left(\frac{3\pi}{2}\left(i-\frac{1}{4}\right)\right)^{2/3}$ and $\alpha_j = \left(2m^*_jeF/\hbar^2\right)^{1/3}$.

Figure \ref{n3d} shows the $n_{3d}$ electron density as a function of $z$. Here, the the Fermi energy $E_F(V)$ is determined self-consistently by calculating 
\be
n_{2d}(V) = \int n_{3d}(z) dz=\int_{0}^{E_F(V)}N_{2d}(E,V)dE
\ee
where $N_{2d}(E,V)$ is the density of states
\be
N_{2d}(E,V)= \sum_{j=l,h}\left(\frac{g_j m_j^*}{2\pi \hbar^2}\sum_i\Theta(\Delta E_{i,j})\right),
\ee
and compare to the value from Eq. \ref{n2d}. We can also determine the average $n_{3d}$ by $\langle n_{3d}\rangle$ = $\int n_{3d}^2 dz$/$\int n_{3d} dz$ and the average thickness $d_{av}=\int n_{3d}zdz/\int n_{3d}dz$. To compare the DOS to the measured capacitance, we consider
\be
\begin{array}{c}
\displaystyle \frac{\delta n_{2d}(V)}{\delta V} = N_{2d}(E_F(V),V)\frac{\delta E_F(V)}{\delta V} \\ \\ \displaystyle + \int_0^{E_f(V)} \frac{\delta}{\delta V}N_{2d}(E,V)dE
\end{array}
\ee
Here, it is shown that as the gate potential is increased, the carriers in the system are increased at the interface. This increases the overall density of carries and shortens the interfacial depth. By using the above determined Fermi energy $E_F(V)$,
the density of states in $2d$ may be related to the derivative of $n_{2d}(V)$
with respect to $V$ and hence to the capacitance as
\be
\begin{array}{c}
\displaystyle N_{2d}(E_{F}(V),V)= \\ \\ \displaystyle \left[\frac{\partial n_{2d}}{\partial V} - \int_0^{E_f(V)} \frac{\delta}{\delta V}N_{2d}(E,V)dE \right]\frac{\partial V}{\partial E_{F}(V)}= \\ \\ \displaystyle
\left[\frac{1}{eS}C(V) - \int_0^{E_f(V)} \frac{\delta}{\delta V}N_{2d}(E,V)dE \right]\frac{\partial V}{\partial E_{F}(V)}
\end{array}
\ee

The inset of Fig. \ref{n3d} shows $E_F$ as a function of $V$. Using a quadratic polynomial fit to the data, we find that $E_F (meV)$ = 4.30+ 0.0410$V$-5.05x10$^{-5}$$V^2$. Therefore, we can used the derivative to find an approximation to $\partial V$ / $\partial E_{F}(V)$. The integral over $N_{2d}(E,V)$ is dependent on the electric-field dependence of the effective masses and $E_{i,j}$. However, it is assumed that this contribution is small within the confines of the interface.

\begin{figure}
\centering
\includegraphics[width=3.75in]{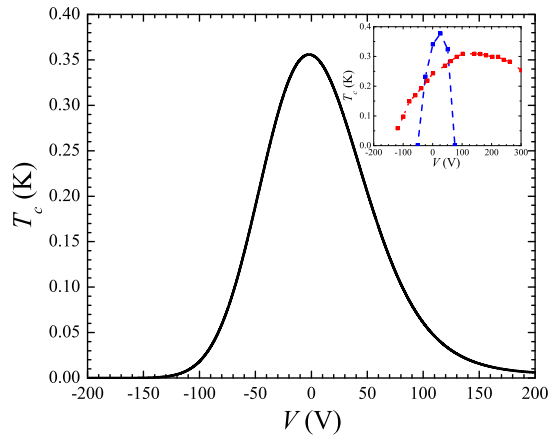}
\caption{Calculated superconducting transition temperature $T_c$ as a function of electric potential for the LAO/STO interface. Here, it is shown that due to the increase in carrier density, there is a distinct change in $T_c$. Eventually, at some critical potential, the Coulomb repulsion takes over and $T_c$ is reduced. Inset shows the experimentally observed change in $T_c$ from Ref. [\onlinecite{bell:09}] (dashed blue) and Ref. [\onlinecite{cavi:08}] (dotted red). The difference in peak width and level can vary depending on the pair interaction's dependence on $V$ as well as the exact determination of $\partial E_{F}(V)$/ $\partial V$.}
\label{Tc}
\end{figure}

\section{Effect on Superconducting transition temperature $T_c$}

To explore the consequences of the change in carrier density induced by
an applied electric field on the superconducting transition temperature. We
exploit the fact that the change in the DOS will modify the
(dimensionless) pair coupling. We conjecture that the pairing is similar to
the one observed in bulk doped STO.\cite{scho:64} There weak
coupling BCS theory provides a quantitative description of the observed
transition temperature as a function of doping. Within weak coupling theory,
the critical temperature $T_{c}$ below which a material becomes
superconducting is given by 
\be
T_c = T_D e^{-\frac{1}{\lambda_p N(E_{F})}},
\ee
where $T_{D}$ is the Debye temperature (513 K for LAO/STO multilayers),  $\lambda_p$ is the effective interaction , and $N(E_{F})$ is the electronic
density of states at the Fermi level in $d=3$ dimensions. In the case of
bulk doped STO, a detailed theoretical study of the transition  temperature
in the framework of Eliashberg's theory has been performed Ref. [\onlinecite{koo:67}]. The conclusion was that inter- and intra-valley
electron-phonon processes as well as Coulomb repulsion contribute to
determining $T_{c}$. These calculations accounted quantitatively for the
observed maximum of  $T_{c}$ as a function of carrier density. The initial
increase of $T_{c}$ at low density is connected with the increase in the
density of states, while the decrease beyond the maximum may be traced to
the weakening of the attraction caused by enhanced screening. Comparing the
values of density $n_{3d}$ found above with the density at the maximum of $T_{c}$ in Ref. [\onlinecite{koo:67}], $n_{3d}^{m}=10^{20}{\rm cm}^{-3}$ we conclude that our systems
are on the low density side, where density dependent screening effects of
the pair interaction are less relevant. We therefore assume that the pair
interaction $\lambda_p$ may be taken as a constant, which we deduce from the
dimensionless coupling, $\lambda_p N(E_{F})=0.13$ and the measured DOS $N(E_{F})=2\times 10^{47}$J$^{-1}$m$^{-3}$ at density $n_{3d}^{m}$ as $\lambda_p
\approx 6.5\times 10^{-49}$Jm$^{3}$. However, the pair interaction may also have a dependence on the applied potential.As shown above the $2d$ electronic density
of states can be directly measured through the capacitance, and the dependence of the Fermi energy on applied voltage, while the latter can be inferred from the observed density-voltage relation. Therefore, the 
$3d$ DOS can be approximately
expressed as 
\be
N(E_{F}(V),V)=\frac{1}{S ed_{av}\gamma}C(V)\frac{\partial V}{\partial E_{F}(V)},
\ee
where $d_{av}$ is the average thickness of the electron layer. There is a second term that we discuss in the supplementary material, but as discussed above, this term is small and can be ignored. Hence, a change in
the capacitance with applied electric field provides a direct effect on $T_{c}$ within this system. Figure \ref{Tc} shows the calculated $T_{c}$ as a
function of $V$ for the LAO/STO heterostructure, where $\gamma \approx$ 0.0045 has
been adjusted to place the peak value of $T_{c}$ at the experimentally
determined 0.37 K. This details the subtle significance of a gate voltage on
the superconducting state. Inset of Fig. \ref{Tc} details and compares the experimentally observed change in $T_c$ from Ref. [\onlinecite{bell:09}] (dashed blue) and Ref. [\onlinecite{cavi:08}] (dotted red). 


The observed transition into the superconductive state is most likely of the Kosterlitz-Thouless-Berezinski (KTB) type. In general the transition temperature $T_{KTB}$ is related to the bulk transition temperature $T_{c}$ by \cite{Beas:79}
\begin{equation}
\frac{T_{KTB}}{T_{c}}\Big/f\left(\frac{T_{KTB}}{T_{c}}\right)=2.18\frac{R_{Q}}{R_{sheet}}
\end{equation}%
where $R_{sheet}$  is the sheet resistance and $R_{Q}=\hbar/e^{2}\approx4.12k\Omega$ is the resistance quantum. Here the function $f$ is defined in terms of the gap parameter $\Delta (T)$ as
\begin{equation}
f\left(\frac{T}{T_{c}}\right)=\frac{\Delta (T)}{\Delta (0)}\tanh \left[\frac{\Delta (T)}{2T}\right]
\end{equation}%
In the limit of weak disorder, $T_{KTB}\approx T_{c}$. At strong disorder, when  $R_{sheet}>>2R_{Q}$ one gets a suppression of the KTB transition temperature as $T_{KTB}=2.18T_{c}R_{Q}/R_{sheet}$.

\section{Conclusion}

In conclusion, we examined the effect of an electric potential on the carrier densities and superconducting transition temperature for the conducting LAO/STO interfaces. We find that the carrier density and superconducting temperature can be significantly modified by an
external electric field. Recently, there have been experimental observations that
have explored the relationship between an applied potential on STO
heterostructures and the increased carrier density or capacitance. Here, we
take this relationship further and assume it is the field effect
that, through nonlinear capacitance, changes carrier density. That in turn
affects the superconducting coupling and ultimately $T_{c}$. We present a mechanism of this phenomenon assuming a conventional pairing state at
the interface, which shows a possible route to tunable material
properties and suggests an exciting perspective on the way to tunable
superconductivity.

\section{Acknowledgements}

We would like to acknowledgement helpful discussions with Q. Jia, D. Yarotski, Y. Liu,  J-X. Zhu, S. A. Trugman, and Th. Kopp.
This work was supported, in part, by the Center for Integrated Nanotechnologies, a U.S. Department of Energy, Office of Basic Energy Sciences user facility and in part by the LDRD and, in part, by UCOP TR-027.  Los Alamos National Laboratory, an affirmative action equal opportunity employer, is operated by Los Alamos National Security, LLC, for the National Nuclear Security Administration of the U.S. Department of Energy under contract DE-AC52-06NA25396. PW gratefully
acknowledges the award of a Carl Schurz Memorial professorship at the
University of Wisconsin, Madison.


\begin{thebibliography}{har}


%
%
%
%
%
%
%
%
%
%
%
%
%
%
%
%
%
%
%
%
%
%
%
%
%
%
%
%
%
%
%
%
%
%




\bibitem{bard:57}
J. Bardeen, L.H. Cooper, and J.R. Schrieffer,Phys. Rev. {\bf 108}, 1175 (1957).

\bibitem{park:69}
R.D. Parks, {\it Superconductivity} (Marcel Dekker, New York, 1969).

\bibitem{ande:97}
P. W. Anderson, {\it Basic Notions of Condensed Matter Physics} (Perseus, Cambridge, 1997).

\bibitem{kett:99}
J. B. Ketterson and S. N. Song, {\it Superconductivity} (Cambridge University Press, New York, 1999).

\bibitem{okam:04}
S. Okamoto and A. J. Millis, Nature {\bf 428}, 630 (2004).

\bibitem{taka:06}
K.S. Takahashi, M. Gabay, D. Jaccard, K. Shibuya, T. Ohnishi, M. Lippmaa, and J.-M. Triscone, Nature {\bf 441}, 195 (2006).

\bibitem{hebe:09}
J. Heber, Nature {\bf 459}, 28 (2009).

\bibitem{hwan:06}
H.Y. Hwang, Mater. Res. Soc. Bull. {\bf 31}, 28 (2006).

\bibitem{ohto:02}
A. Ohtomo, D. A. Muller, J. L. Grazul, and H.Y. Hwang, Nature {\bf 419}, 378 (2002).

\bibitem{ohto:04}
A. Ohtomo and H.Y. Hwang, Nature {\bf 427}, 423 (2004).

\bibitem{reyr:07}
N. Reyren, S. Thiel, A. D. Caviglia, L. Fitting Kourkoutis, G. Hammerl, C. Richter, C. W. Schneider, T. Kopp, A.-S. R\"uetschi, D. Jaccard, M. Gabay, D. A. Muller, J.-M. Triscone, and J. Mannhart, Science {\bf 317}, 1196 (2007).

\bibitem{siem:07}
W. Siemons, G. Koster, H. Yamamoto, W. A. Harrison, G. Lucovsky, T. H. Geballe, D. H. A. Blank, and M. R. Beasley, Phys. Rev. Lett. {\bf 98}, 196802 (2007).

\bibitem{naka:06}
N. Nakagawa, H. Y. Hwang, and D. A. Muller, Nature Materials {\bf 5}, 204 (2006).

\bibitem{waru:09}
M. P. Warusawithana, C. Cen, C. R. Sleasman, J. C. Woicik, Y. Li, L. Fitting Kourkoutis, J. A. Klug, H. Li, P. Ryan, L.-P. Wang, M. Bedzyk, D. A. Muller, L.-Q. Chen, J. Levy, and D. G. Schlom, Science {\bf 324}, 367 (2009)

\bibitem{cavi:08}
A. D. Caviglia, S. Gariglio, N. Reyren, D. Jaccard, T. Schneider, M. Gabay, S. Thiel, G. Hammerl, J. Mannhart, and J.-M. Triscone, Nature {bf 456}, 624 (2008).

\bibitem{bell:09}
C. Bell, S. Harashima, Y. Kozuka, M. Kim, B. G. Kim, Y. Hikita, and H. Y. Hwang, Phys. Rev. Lett. {\bf 103}, 226802 (2009).

\bibitem{scho:64}
J. F. Schooley, W. R. Hosler, and M. L. Cohen, Phys. Rev. Lett. {\bf 12} 474 (1964).

\bibitem{ueno:08}
K. Ueno, S. Nakamura, H. Shimotani, A. Ohtomo, N. Kimura, T. Nojima, H. Aoki, Y. Iwasa, and M. Kawasaki, Nature Materials {\bf 7}, 855 (2008).

\bibitem{koer:05}
V. Koerting, Q. Yuan, P. J. Hirschfeld, T. Kopp, and J. Mannhardt, Phys. Rev. B {\bf 71}, 104510 (2005).

\bibitem{step:11}
C. Stephanos, T. Kopp, J. Mannhart, and P. J. Hirschfeld, arXiv:1108.1942 (2011).

\bibitem{bert:11}
J. A. Bert, B. Kalisky, C. Bell, M. Kim, Y. Hikita, H. Y. Hwang, and K. A. Moler, Nature Physics {\bf 7}, 767 (2011).

\bibitem{brink:07}
A. Brinkman, M. Huijben, M. van Zalk, J. Huijben, U. Zeitler, J. C. Maan, W. G. van der Wiel, G. Rijnders, D. H. A. Blank, and H. Hilgenkamp, Nature Materials {\bf 6}, 493 (2007).

\bibitem{li:11}
L. Li, C. Richter, J. Mannhardt, and R. C. Ashoori,  Nature Physics {\bf 7}, 762 (2011).

\bibitem{millis:11}
A. J. Millis, Nature Physics {\bf 7}, 749 (2011).

\bibitem{thiel:06}
S. Thiel, G. Hammerl, A. Schmehl, C. W. Schneider, and J. Mannhardt, Science {\bf 313}, 1942 (2006).

\bibitem{sing:09}
M. Sing, G. Berner, K. Go\ss, A. M\"uller, A. Ruff, A. Wetscherek, S. Thiel, J. Mannhart, S. A. Pauli, C. W. Schneider, P. R. Willmott, M. Gorgoi, F. Sch\"afers4, and R. Claessen, Phys. Rev. Lett. {\bf 102}, 176805 (2009).

\bibitem{hara:11}
J.T. Haraldsen, S.A. Trugman, A.V. Balatsky, Phys. Rev. B {\bf 84}, 020103(R) (2011).

\bibitem{linearnote}
Given the electro-static potential interaction $\int \rho_z |\psi_{z^{\prime}}|^2 K(z,z^{\prime}) dz dz^{\prime}$, where $\rho_z$ is the charge density and is proportional to $c_1P_0$ and $K(z,z^{\prime})$ is the electro-static kernel. Here, $\int |\psi_{z^{\prime}}|^2 K(z,z^{\prime}) dz^{\prime}$ is proportional to $\kappa |\psi_0|^2$. Therefore, the linear interaction term at the interface is derived as $c_1 \kappa P_0  |\psi_0|^2$ or $\lambda P_0  |\psi_0|^2$.

\bibitem{serq:01}
A. Serquis, Y. T. Zhu, E. J. Peterson, J. Y. Coulter, D. E. Peterson, and F. M. Mueller, Appl. Phys. Lett. {\bf 79}, 4399 (2001).


\bibitem{sole:02}
J. M. Soler, E. Artacho, J. D. Gale, A. Garc\'ia, J. Junquera, P. Ordej\'on, and D. S\'anchez-Portal, J. Phys.: Condens. Matter {\bf 14}, 2745 (2002).

\bibitem{quantum}
Calculations were performed using the Atomistix Toolkit by QuantumWise. ÒATK version 11.8.b1Ó, QuantumWise A/S (www.quantumwise.com).

\bibitem{ster:72}
F. Stern, Phys. Rev. B {\bf 5}, 4891 (1972).

\bibitem{koo:67}
C. S. Koonce, M. L. Cohen, J. F. Schooley, W. R. Hosler, and E. R. Pfeiffer, Phys. Rev. {\bf 163}, 380 (1967).

\bibitem{Beas:79}
M. R. Beasley, J. P. Mooij, and T. P. Orlando, Phys. Rev. Lett. {\bf 42}, 1165 (1979).


\end{thebibliography}
\end{document}